\documentclass[10pt,letterpaper]{article}
\usepackage{opex3}
\usepackage{epsfig,amsmath,graphicx,amssymb,color,cite}
\usepackage{bm}
\begin{document}
\renewcommand{\pl}{\partial}
\newcommand{\be}[0]{\begin{equation}}
\newcommand{\ee}[0]{\end{equation}}
\newcommand{\f}[2]{\frac{#1}{#2}}
\def\bea{\begin{eqnarray}}
\def\eea{\end{eqnarray}}
\def\bes{\begin{subequations}}
\def\ees{\end{subequations}}
\title{PT symmetry via electromagnetically induced transparency}
\author{Hui-jun Li,$^{1,2,\ast}$ Jian-peng Dou,$^{1}$ and Guoxiang Huang$^{2,3,4}$}

\address{$^1$Institute of Nonlinear Physics and Department of
              Physics, Zhejiang Normal University, Jinhua, 321004 Zhejiang, China\\
              $^2$State Key Laboratory of Precision Spectroscopy, East China Normal University, 200062 Shanghai, China\\
              $^3$Department of Physics, East China Normal University, 200062 Shanghai, China\\
              $^4$gxhuang@ecnu.edu.cn}

\email{$^{\ast}$ hjli@zjnu.cn} 



\begin{abstract} We propose a scheme to realize  parity-time (PT) symmetry
via electromagnetically induced transparency (EIT). The system we consider
is an ensemble of cold four-level atoms with an EIT core.
We show that the cross-phase modulation contributed by an assisted field,
the optical lattice potential provided by a far-detuned laser field, and
the optical gain resulted from an incoherent pumping can be used to construct
a PT-symmetric complex optical potential
for probe field propagation in a controllable way. Comparing with previous study,
the present scheme uses only a single atomic species and hence is easy for the
physical realization of PT-symmetric Hamiltonian via
atomic coherence.\end{abstract}



\section{Introduction}\label{sec1}

%

In recent years, a lot of efforts have been made on a class of non-Hermitian Hamiltonian with
parity-time (PT) symmetry, which in a definite range of system parameters
may have an entirely real spectrum~\cite{ben0,mak0}. PT symmetry requires that
the real (imaginary) part of the complex potential in the Hamiltonian is an even (odd)
function of space, i.e.  $V({\bf r})= V^*({-\bf r})$.
Even though the Hermiticity of quantum observables has been widely accepted, there is still
great interest in  PT symmetry because of the motivation for constructing a
framework to extend or replace the Hermiticity of the Hamiltonian in ordinary quantum mechanics.
The concept of PT symmetry has also stimulated many other studies, such as quantum
field theory~\cite{ben1}, non-Hermitian Anderson models~\cite{gol}, and open quantum systems~\cite{rot},
and so on.

Although a large amount of theoretical works exist, the
experimental realization of PT-symmetric Hamiltonian in the fields
mentioned above was never achieved. Recently, much attention has
been paid to various optical systems where PT-symmetric
Hamiltonians can be realized experimentally by balancing optical
gain and loss~\cite{rusch,elg,mak,mus}. In optics, PT symmetry is
equivalent to demand a complex refractive index with the property
$n({\bf r})= n^*({-\bf r})$. Such refractive index has been
realized experimentally using two-wave mixing in an Fe-doped
LiNbO$_3$ substrate~\cite{rut}. The optical realization of PT
symmetry has motivated various designs of PT-synthetic optical
materials exhibiting many intriguing features, including
non-reciprocal or unidirectional reflectionless wave
propagation~\cite{rut,kul,lin1,feng1}, coherent perfect
absorber~\cite{long,chong}, giant wave amplification~\cite{kon},
etc. Experimental realization of PT symmetry using
plasmonics~\cite{ben}, synthetic lattices~\cite{reg}, and LRC
circuits~\cite{schin} were also reported.

In a recent work Hang {\it et al.}~\cite{hang} proposed a double
Raman resonance scheme to realize PT symmetry by using a
two-species atomic gas with $\Lambda$-type level configuration.
This scheme is quite different from those based on solid systems
mentioned
above~\cite{rusch,rut,kul,lin1,feng1,long,chong,kon,ben,reg,schin},
and possesses many attractive features. For instance, the
PT-symmetric refractive index obtained in~\cite{hang} is valid in
the whole space; furthermore, the refractive index can be actively
controlled and precisely manipulated by changing the system
parameters {\it in situ}.

In the present article, we suggest a new scheme to realize the PT
symmetry in a lifetime-broadened atomic gas based on the mechanism
of electromagnetically induced transparency (EIT), a typical and
important quantum interference phenomenon widely occurring in
coherent atomic systems~\cite{fle}. 
Different from the two-species, double Raman resonance scheme
proposed in~\cite{hang}, the scheme we suggest here is a
single-species, EIT one. And due to the complexity of the
susceptibility~\cite{hang}, it is difficult to design some PT
potentials we wish, however, in our scheme, we can design many
different periodic potentials and non-periodic potentials in light
of our will, and the size of potential can also be  adjusted
conveniently. Especially, compared with the traditional idea that
PT symmetric potential must be combined by the gain and loss
parts, we utilize the atomic decay rate to design the imaginary
part of PT potential, and use the giant cross-phase modulation
(CPM) effect~\cite{fle,schm} of the resonant EIT system to realize
the real part. We shall show that the cross-phase modulation
contributed by the assisted field, the optical lattice potential
provided by a far-detuned laser field, and the optical gain
resulted from an incoherent pump can be used to construct a
complex optical potential with PT symmetry for probe field
propagation in a controllable way. The present scheme uses a
single atomic species only and hence is simple for physical
realization.

The rest of the article is arranged as follows. In the next section, a description of our scheme
and basic equations for the motion of atoms and light field are presented. In Sec.~III,
the envelope equation of the probe field and its realization of PT symmetry
are derived and discussed. The final section is the summary of our main results.

\section{Model and equations of motion}
\label{sec2}

\subsection{Model}\label{sec21}

\begin{figure}
\centering
\includegraphics[scale=0.6]{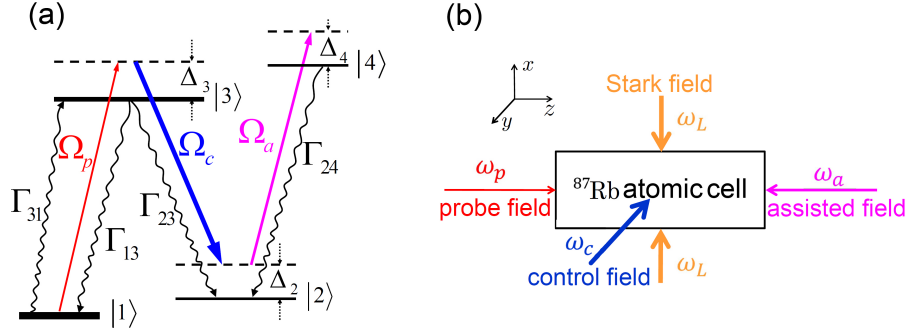}
\caption{\footnotesize (a) Energy-level diagram and excitation
scheme used for obtaining a PT symmetric model. (b) Possible
experimental arrangement. All the notation are defined in the
text.} \label{fig1}
\end{figure}

The system under consideration is a cold, lifetime-broadened
$^{87}$Rb atomic gas with N-type level configuration; see
Figure~\ref{fig1}.
%
The levels of the system are taken from the D$_{1}$ line of
$^{87}$Rb atoms, with $|1\rangle=|5 S_{1/2}, F=1\rangle$,
$|2\rangle=|5 S_{1/2}, F=2\rangle$, $|3\rangle=|5 P_{1/2},
F=1\rangle$, and $|4\rangle=|5 P_{1/2}, F=2 \rangle$. A weak probe
field ${\bf E}_{p}={\bf e}_x {\cal E}_{p}(z,t)\exp{[i(k_{p}
z-\omega_{p}t)]}+{\rm c.c.}$ and a strong control field ${\bf
E}_{c}={\bf e}_{x}{\cal E}_{c}\exp{[i(-k_{c} y-\omega_{c}t)]}+{\rm
c.c.}$ interact resonantly with levels
$|1\rangle\rightarrow|3\rangle$ and $|2\rangle\rightarrow
|3\rangle$, respectively.  Here ${\bf e}_{j}$ and $k_{j}$ (${\cal
E}_{j}$) are respectively the polarization unit vector in the
$j$th direction and the wave number (envelope) of the $j$th field.
The levels $|l\rangle$ ($l=1,2,3$) together with ${\bf E}_{p}$ and
${\bf E}_{c}$ constitute a well-known $\Lambda$-type EIT core.

Furthermore, we assume an assisted filed
\be \label{assist}
{\bf E}_{a}={\bf e}_{y}{\cal E}_{a}(x) \,\exp {[i(-k_{a} z-\omega_{a}t)]}+{\rm c.c.}
\ee
is coupled to the levels $|2\rangle\rightarrow|4\rangle$, where
${\cal E}_{a}(x)$ is field-distribution function in transverse direction.
The assisted filed ${\bf E}_{a}$, when assumed to be weak
(satisfying ${\cal E}_p \le {\cal E}_a \ll {\cal E}_c$), will contribute a
CPM effect to the probe field ${\bf E}_{p}$.
Note that the levels
$|l\rangle$ ($l=1,2,3,4$) together with ${\bf E}_{p}$, ${\bf E}_{c}$, and  ${\bf E}_{a}$
form a N-type system, which was considered firstly by Schmidt Imamo\v{g}lu~\cite{schm}
for obtaining giant CPM via EIT.

In addition, we assume there is another far-detuned (Stark) optical lattice field
\be\label{stark}
{\bf E}_{\rm Stark}=\textbf{e}_{y} \sqrt{2} E_s(x) \cos (\omega_L t)
\ee
is applied to the system, where $E_s(x)$ and $\omega_L$ are respectively the
field-distribution function and angular frequency.
Due to the existence of ${\bf E}_{\rm Stark}$, a small and
$x$-dependent Stark shift of level $E_j$ to the state $|j\rangle$
occurs, i.e., $E_j\rightarrow E_j+\Delta E_j$ with $\Delta
E_j=-\frac{1}{2}\alpha_j \left<{\bf E}_{\rm
Stark}^2\right>_t=-\frac{1}{2}\alpha_j |E_{s}(x)|^{2}$, here
$\alpha_j$ is the scalar polarizability of the level $|j\rangle$,
and $\langle \cdots\rangle_t$ denotes the time average in an
oscillating cycle. The explicit forms of ${\cal E}_{a}(x)$ and
$E_s(x)$ in (\ref{assist}) and (\ref{stark}) will be chosen later
on according to the requirement of  PT symmetry (see Sec.~3.2).

As will be shown below, the CPM effect contributed by the assisted
field ${\bf E}_{a}$ given by (\ref{assist}) and the Stark shift
contributed by the far-detuned Stark field ${\bf E}_{\rm Stark}$
given by (\ref{stark}) will provide periodic complex refractive
index to the evolution of  probe-filed envelope. However, they are
still not enough to obtain a refractive index with PT symmetry
since a gain to the probe field is needed. Therefore, we introduce
an incoherent optical pumping which can pump atoms from the
ground-state level $|1\rangle$ to the excited-state level
$|3\rangle$ with the pumping rate $\Gamma_{31}$ [see equations
(18a) and (18c) in Appendix]. Such optical pumping can be realized
by many techniques, such as intense atomic resonance lines emitted
from hollow-cathode lamps or from microwave discharge
lamps~\cite{dem}.

In Fig.~1(a),  $\Gamma_{13}$, $\Gamma_{23}$, and $\Gamma_{24}$ are
spontaneous emission rates denoting the population decays
respectively from $|3\rangle$ to $|1\rangle$,  $|3\rangle$ to
$|2\rangle$, and  $|4\rangle$ to  $|2\rangle$; $\Omega_p = ({\bf
e}_{x}\cdot {\bf p}_{13}){\cal E}_{p}/\hbar$, $\Omega_c =({\bf
e}_{x}\cdot{\bf p}_{23}){\cal E}_{c}/\hbar$, and $\Omega_{a}=({\bf
e}_{y}\cdot{\bf p}_{24}){\cal E}_{a}/\hbar$ are respectively the
half Rabi frequencies of the probe, control, and assisted fields,
here ${\bf p}_{ij}$ signifies the electric dipole matrix element
of the transition from state $|i\rangle $ to $|j\rangle$,
$\Delta_3$,  $\Delta_2$, and $\Delta_4$ are respectively one-,
two-, and three-photon detunings in relevant transitions.
Fig.~1(b) shows a possible experimental arrangement.

\subsection{Maxwell-Bloch equations}

Under electric-dipole and rotating-wave approximations, the Hamiltonian
of the system in interaction picture reads
$ \hat{H}_{\rm int}=-\hbar\sum_{j=1}^{4}\Delta_{j}^{\prime}|j\rangle\langle
j|-\hbar(\Omega_{p}|3\rangle\langle 1|+\Omega_{c}|3\rangle\langle 2|+\Omega_{a}|4\rangle\langle 2|+{\rm h. c.}),$
where ${\rm h. c.}$ denotes Hermitian conjugate, and
\be \label{Det}
\Delta_{j}^{\prime}=\Delta_{j}+\frac{\alpha_{j}}{2\hbar}|E_{s}(x)|^{2}.
\ee
The motion of atoms interacting with the light fields is described by the
Bloch equation
\be\label{bloche}
\f{\pl \sigma}{\pl t}=-\f{i}{\hbar}\left[\hat{H}_{\rm
int},\sigma\right]-\Gamma\, \sigma,
\ee
where $\sigma_{jl}$ is the density-matrix elements in the
interaction picture, $\Gamma $ is a $4\times 4$ relaxation matrix.
Explicit expressions of Eq.~(\ref{bloche}) are presented in
Appendix, in which an incoherent optical pumping (represented by
$\Gamma_{31}$) from the level $|1\rangle$ to the level $|3\rangle$
is introduced [see equations~(\ref{den1}) and (\ref{den3})].

Under a slowly varying envelope approximation,  Maxwell equation of the probe
field is reduced to
\be\label{max}
i\left(\f{\pl}{\pl z}+\f{1}{c} \f{\pl}{\pl t }\right)\Omega_{p}+
\f{c}{2\omega_{p}}\f{\pl
^{2}\Omega_{p}}{\pl x^{2}}+\kappa_{13}\sigma_{31}=0,
\ee
where
$\kappa_{13}=N\omega_{p}|\textbf{e}_{x}\cdot\textbf{p}_{13}|^2/(2\epsilon_{0}\hbar
c)$ with $N$ being the atomic concentration. Note that, for
simplicity, we have assumed $\Omega_p$ is independent on $y$,
which is valid only for the probe beam having a large width in the
$y$-direction so that the diffraction term $\partial^2 \Omega_p/\partial y^2$
can be neglected; in addition, we have also assumed that the dynamics
of $\Omega_a$ is negligible during probe-field evolution, which is
a reasonable approximation because the assisted field couples to
the levels $|2\rangle$ and $|4\rangle$ that have always vanishing
population due to the EIT effect induced by the strong control
field.

\section{Realization of PT symmetric potential}\label{sec3}

\subsection{Equation of the probe-field envelope}\label{sec3a}

The Maxwell equation~(\ref{max}) governs the propagation of the
probe field. To solve it one must know $\sigma_{31}$, which is
controlled by the Bloch equation~(\ref{bloche}) and hence coupled
to $\Omega_{p}$. For simplicity,  we assume $\Omega_{p}$ has a
large time duration $\tau_0$ so that $\Gamma_{31}\tau_0>>1$. In this case
a continuous-wave approximation can be
taken. As a result, the time derivatives in the Maxwell-Bloch (MB)
equations~(\ref{bloche}) and (\ref{max}) (i.e. the dispersion
effect of the probe field) can be neglected, and only the
diffraction effect of the probe field in $x$ direction is
considered. In addition, because the probe field is weak, a
perturbation expansion can be used for solving coupled
equations~(\ref{bloche}) and (\ref{max})
analytically~\cite{hua,li}.

We take the expansion
$\sigma_{ij}=\sigma_{ij}^{(0)}+\epsilon
\sigma_{ij}^{(1)}+\epsilon^{2}
\sigma_{ij}^{(2)}+\epsilon^{3}\sigma_{ij}^{(3)}+\cdots$,
$\Omega_{p}=\epsilon \Omega_{p}^{(1)}
+\epsilon^{3}\Omega_{p}^{(3)} +\cdots$. Here  $\epsilon$ is a
small parameter characterizing the typical amplitude of the probe
field (i.e  $\Omega_{p,{\rm max}}/\Omega_{c}$). Substituting such
expansion to equations~(\ref{bloche}) and (\ref{max}), we obtain a
series of linear but inhomogeneous equations for
$\sigma_{ij}^{(l)}$ and $\Omega_{p}^{(l)}$ ($l=1,2,3,...$) that
can be solved order by order. To get a divergence-free
perturbation expansion, $\sigma_{ij}^{(l)}$ and $\Omega_{p}^{(l)}$
are considered as functions of the multiple scale variables
$z_{l}=\epsilon^{l}z$ ($l=0,\,2$) and $x_{1}=\epsilon
x$~\cite{hua,li}. In addition, we assume $\Omega_{a}=\epsilon
\Omega_{a}^{(1)}(x_1)$, $E_{s}=\epsilon E_{s}^{(1)}(x_1)$. Thus we
have $d_{ij}=d_{ij}^{(0)}+\epsilon^{2}d_{ij}^{(2)}$ with
$d_{ij}^{(0)}=\Delta_{i}-\Delta_{j}+i\gamma_{ij}$ and
$d_{ij}^{(2)}=[(\alpha_{i}-\alpha_{j})/(2\hbar)]
|E_{s}^{(1)}|^{2}$.

At ${\cal O}(1)$-order, we obtain non-zero density-matrix elements
$\sigma_{11}^{(0)}=1-(2-X_{1})X_{2}$,
$\sigma_{22}^{(0)}=(1-X_{1})X_{2},$ $\sigma_{33}^{(0)}=X_{2}$,
$\sigma_{32}^{(0)}=[\Omega_{c}^{\ast}/(d_{32}^{(0)})^{\ast}]X_{1}X_{2}$,
with $X_{1}=\Gamma_{23}/[2{\rm Im}(|\Omega_{c}/d_{32}^{(0)}|^{2})]$
and $X_{2}=\Gamma_{31}/[\Gamma_{13}+\Gamma_{31}(2-X_{1})]$. It is
the base state solution of the MB equations (i.e., the solution for $\Omega_p=\Omega_a=0$).
We see that due to the existence of the incoherent optical pumping (i.e., $\Gamma_{31}\neq 0$)
there are populations in the states $|1\rangle$, $|2\rangle$, and $|3\rangle$. Because
$\Gamma_{31}$ takes the order of MHz in our model, the populations in $|2\rangle$ and $|3\rangle$ are
small. In particular, $\sigma_{22}^{(0)}=\sigma_{33}^{(0)}=0$, $\sigma_{11}^{(0)}=1$
when $\Gamma_{31}= 0$.

At ${\cal O}(\epsilon)$-order, the solution is given by
\bes
\bea
& & \label{op1} \Omega_{p}^{(1)}=F\,e^{i K z_{0}},\\
& & \label{s211}
\sigma_{21}^{(1)}=\f{\Omega_{c}^{\ast}(\sigma_{33}^{(0)}-\sigma_{11}^{(0)})
-d_{31}^{(0)}\sigma_{23}^{(0)}}{D_{1}}F
e^{i K z_{0}}\equiv \alpha_{21}^{(1)}F e^{i K z_{0}},\\
& & \label{s311} \sigma_{31}^{ (1)}=\f{K}{\kappa_{13}}F
e^{i K z_{0}}\equiv\alpha_{31}^{(1)}F e^{i K z_{0}},\\
& & \label{s421}
\sigma_{42}^{(1)}=\f{d_{43}^{(0)}\sigma_{22}^{(0)}+\Omega_{c}
\sigma_{23}^{(0)}}{D_{2}}\Omega_{a}^{(1)}\equiv\alpha_{42}^{(1)}\Omega_{a}^{(1)},\\
& &\label{s431}
\sigma_{43}^{(1)}=\f{\Omega_{c}^{\ast}\sigma_{22}^{(0)}
+d_{42}^{(0)}\sigma_{23}^{(0)}}{D_{2}}\Omega_{a}^{(1)}
\equiv\alpha_{43}^{(1)}\Omega_{a}^{(1)},
\eea
\ees
with other $\sigma_{jl}^{(1)}=0$. Here $F$ is yet to be determined envelope function,
$D_{1}=|\Omega_{c}|^{2}- d_{21}^{(0)}d_{31}^{(0)}$,
$D_{2}=|\Omega_{c}|^{2}-d_{42}^{(0)}d_{43}^{(0)}$, and
\be\label{K}
K=\kappa_{13} \f{d_{21}^{(0)}(\sigma_{11}^{(0)}-\sigma_{33}^{(0)})+\Omega_{c}\sigma_{23}
^{(0)}}{D_{1}}.
\ee
Obviously, in the linear case $\Omega_{p}\propto e^{i K z},$ and
$K$ is complex. Thus $K$, particularly its imaginary part,
controls the behavior of the probe-field propagating along $z$.

%
\begin{figure}
\centering
\includegraphics[scale=0.4]{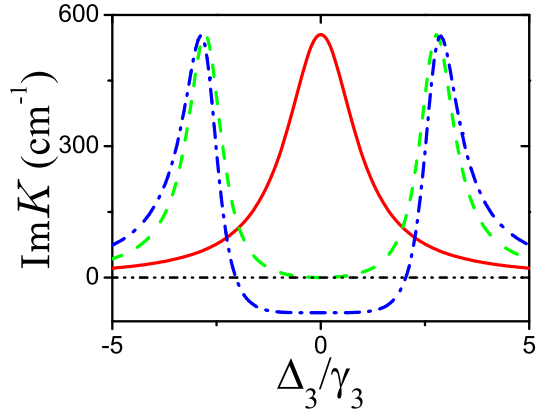}
\caption{\footnotesize The imaginary part Im$K$ of $K$ as a
function of $\Delta_{3}/\gamma_3$ for $\Delta_{2}=\Delta_{3}$.
Solid (red), dashed (green), and dashed-dotted (blue) lines
correspond to $(\Omega_c,\Gamma_{31})=(0,0)$, ($5\times 10^7$ Hz,
0), and ($5\times 10^7$ Hz, $0.7\gamma_{3}$), respectively. For
illustration, the value of dashed-dotted (green) line has been
amplified 7.8 times. } \label{fig2}
\end{figure}
%
Figure~\ref{fig2} shows the imaginary part Im$K$ of $K$ as a
function of $\Delta_{3}/\gamma_3$ for $\Delta_{2}=\Delta_{3}$. The
system parameters used are~\cite{steck}  $\gamma_{1}=\Delta_{1}=0
\,{\rm Hz},\, 2\gamma_{2}=1\times10^{3} \,{\rm
Hz},\,\Gamma_{3}=2\gamma_{3}=36\,{\rm MHz},\,\kappa_{13}=1.0\times
10^{10} \, {\rm cm}^{-1}$ Hz. Solid (red), dashed (green), and
dashed-dotted (blue) lines correspond to
$(\Omega_c,\Gamma_{31})=(0, 0)$, ($5\times 10^7$ Hz, 0), and
($5\times 10^7$ Hz, $0.7\gamma_{3}$), respectively.

From the solid line of Figure~\ref{fig2}, we see that in the
absence of the control field and incoherent pumping (i.e.,
$\Omega_c=\Gamma_{31}=0$), the probe field has a very large
absorption; however, when the incoherent pumping still absent but
$\Omega_c$ takes the value of $5\times 10^7$ Hz, a transparency
window is opened (as shown by the dashed line). This is well-known
EIT quantum interference phenomenon induced by the control
field~\cite{fle}. However, there is still a small absorption
(i.e., Im$K>0$, which can not be seen clearly due to the
resolution of the figure). That is to say, although EIT can
suppress largely the absorption, it can not make the absorption
become zero.

The dashed-dotted line in Fig.~\ref{fig2} is the situation when
the incoherent pumping ($\Gamma_{31}=0.7\gamma_{3}$) is
introduced. One sees that a gain (i.e., negative Im$K$ in the
region near $\Delta_{3}=0$) occurs. Such gain is necessary to get
a PT-symmetric optical potential for the probe-field propagation,
as shown below.

At ${\cal O}(\epsilon^{3})$-order of the perturbation expansion,
we obtain the closed equation for $F$, which can be converted to
the equation for $\Omega_p$:
\be\label{SE0} i\frac{\partial \Omega_{p}}{\partial
z}+\f{c}{2\omega_{p}} \frac{\partial^{2} \Omega_{p}}{\partial
x^{2}}+ \tilde{V}(x)\Omega_{p}=0 \ee
after returning to original variables, with \be \label{Vt}
\tilde{V}(x)=\alpha_{12}\frac{|{\bf e}_y\cdot {\bf p}_{24}|^2}
{\hbar^2} |{\cal E}_{a}(x)|^{2}+\alpha_{13}|E_{s}(x)|^{2}+K, \ee
where $\Omega_{p}=\epsilon F\exp(iKz)$, the coefficients
$\alpha_{12}$ and $\alpha_{13}$ are given in Appendix.

We now make some remarks about the potential $\tilde{V}(x)$ given by Eq.~(\ref{Vt}):

(1). The coefficients $\alpha_{12}$ and $\alpha_{13}$  are {\it
complex}. We stress that the occurrence of a complex  potential
for the evolution of probe-field envelope is a general feature in
the system with resonant interactions. The reason is that, due to
the resonance, the finite lifetime of atomic energy states must be
taken into account. As a result, the variation of the probe-field
wavevector resulted by the external light laser fields (here the
Stark and the assisted fields) are complex. It is just this point
that provides us the possibility to realize a PT  symmetric
potential in our system by using the periodic external laser
fields.

(2). If the incoherent pumping is absent, the probe field has only
absorption but no gain and hence not possible to realize PT
symmetry. With the incoherent pumping present, the parameter $K$
[given by the Eq.~(\ref{K})] in the Eq.~(\ref{Vt}) is complex and
has negative imaginary part in the region near $\Delta_3=0$, which
can be used to suppress an absorption constant (i.e. the term not
dependent on $x$)  appearing in the previous two terms of
$\tilde{V}(x)$.

(3). It is easy to show that if only a single external laser field
(the Stark or the assisted field) is applied, it is impossible to
realize a PT  symmetry. That is why the two separated light fields
(i.e. both the Stark and the assisted fields) have been adopted.
We shall show below that the joint action  between the Stark
field,  the assisted field, and the incoherent pumping can give
PT-symmetric potentials in the system.

The susceptibility of the probe field  is given by $\chi
(x)=2c\tilde{V}(x)/\omega_{p}.$  Because the potential (\ref{Vt})
is a complex function of $x$, which is equivalent to a
space-dependent complex refractive index $n (x)=\sqrt{1+\chi
(x)}\approx 1+c\tilde{V}(x)/\omega_{p}$ for the probe-field
propagation. PT symmetry requires $\tilde{V}^*(-x)=\tilde{V}(x)$,
which is equivalent to the condition $n^*(-x)=n(x)$.

\subsection{The design of PT symmetric potential}

Equation (\ref{SE0}) is a linear Schr\"{o}dinger equation with
the ``external'' potential (\ref{Vt}). To realize a PT-symmetric
model we assume the field-distribution functions in (\ref{assist})
and (\ref{stark}) taking the forms
\bea
& & \label{Ea} {\cal E}_a (x)=E_{a0} [\cos (x/R)+\sin (x/R)],\\
& & \label{Es} E_s(x)=E_{s0} \cos (x/R),
\eea
with $E_{a0}$ and $E_{s0}$ being typical amplitudes and $R^{-1}$
being typical ``optical lattice'' parameter. For convenience of
later discussion, we write Eq.~(\ref{SE0}) into the following
dimensionless form
\be
\label{GNLSE2} i\f{\pl u}{\pl s}+\f{\pl ^{2}u}{\pl
\xi^{2}}+V(\xi)u=0,
\ee
with
\be
\label{ptp}
V(\xi)=(g_{12}+g_{12}\sin{2\xi})+g_{13}\cos^{2}{\xi}+K_{0},
\ee
where $u=\Omega_{p}/U_{0},$ $s=z/L_{\rm diff},$ $\xi=x/R$,
$g_{12}=\alpha_{12}|{\bf e}_y\cdot {\bf p}_{24}|^2E_{a0}^{2}L_{\rm
diff}/\hbar^2,$ $g_{13}=\alpha_{13}E_{s0}^{2}L_{\rm diff}$, and
$K_{0}=K L_{\rm diff}.$ Here, $L_{\rm diff}\equiv
2\omega_{p}R^{2}/c$ is the typical diffraction length and $U_{0}$
denotes the typical Rabi frequency of the probe field.

PT symmetry of Eq.~(\ref{GNLSE2}) requires $V^*(-\xi)=V(\xi)$. In
general, such requirement is difficult to be satisfied because
resonant atomic systems have very significant absorption. However,
in the system suggested here the absorption can be largely
suppressed by the EIT effect induced by the control field. The
remainder small absorption that can not be eliminated by the EIT
effect may be further suppressed by the introduction of the
incoherent optical pumping. If the optical pumping is large
enough, the system can acquire a gain. This point can be
understood from Fig.~\ref{fig2} for the case of
$(\Omega_c,\Gamma_{31})=(5\times 10^7$ Hz, $0.7\gamma_{3}$) where
near the EIT transparency window Im$K$ is negative, which means
that the probe field acquires a gain contributed by the optical
pumping. Such gain can be used to suppress the imaginary parts of
$g_{12}$ and $g_{13}$ through choosing suitable system parameters,
and hence one can realize a PT symmetry of the system.

For a practical example, we select the D$_{1}$ line of $^{87}$Rb
atoms, with the energy levels indicated in the beginning of
Sec.~\ref{sec21}. The system parameters are given by
$2\gamma_{2}=1\times 10^{3}\,\,{\rm Hz}$,
$\Gamma_{3,4}=$$2\gamma_{3,4}=36\,\,{\rm MHz}$, $|{\bf
p}_{24}|=2.54\times 10^{27}$ C cm, $\omega_{p}=2.37\times
10^{15}\,{\rm s}^{-1}$. Other (adjustable) parameters are taken as
$\kappa_{13}=2.06\times 10^{11}\,\,{\rm cm}^{-1} {\rm s}^{-1}$,
$R=2.5\times 10^{-3}\,{\rm cm}$, $\Omega_{c}=4.0\times10^{8}\,{\rm
s}^{-1}$, $\Delta_{2}=-5.0\times10^{5}\,{\rm s}^{-1}$,
$\Delta_{3}=5.0\times10^{8}\,{\rm s}^{-1}$,  and $\Delta_{4}=0$.
Then we have $L_{\rm diff}=1.0$ cm, and
 \bea
 & & \label{ASS} {\cal E}_a (x)=0.1 \,(\cos {\xi}+\sin{\xi})\,\, {\rm V/cm},\\
 & & \label{FD} E_s(x)=4.51\times 10^5 \cos{\xi} \,\,{\rm V/cm},\\
 & & \label{OP} \Gamma_{31}=7.0\times 10^{5}\,\,{\rm Hz}.
 \eea
Based on these data and the assisted laser field (\ref{ASS}), the
far-detuned laser field (\ref{FD}) and the optical pumping
(\ref{OP}), we have $g_{12}=0.01+0.4i$, $g_{13}=1.00+0.03i$, and
$K_{0}=-11.7-0.4i$. Here, the imaginary parts of $g_{12}$ and
$K_{0}$ can be alone controlled  by ${\cal E}_a (x)$ and
$\kappa_{13}$, respectively. As a result, we obtain
\be\label{ptp1}
V(\xi)=-11.7+\cos^{2}{\xi}+0.4i\sin{2\xi}+{\cal O}(10^{-2}).
\ee Equation~(\ref{ptp1}) satisfies the  PT-symmetry requirement
$V^*(-\xi)=V(\xi)$ when exact to the accuracy ${\cal O}(10^{-2})$.
The constant term $-11.7$ in  $V(\xi)$ can be removed by using a
phase transformation $u\rightarrow u\exp (-i 11.7 s)$. Equation
(\ref{ptp1}) is a kind of PT-symmetric periodic potential. In
fact, one can design many different periodic potentials or
non-periodic potentials with PT symmetry in our system by using
different assisted and far-detuned laser fields. Consequently, our
system has obvious advantages for actively designing different
PT-symmetric optical potentials and manipulating them in a
controllable way.

\section{Conclusion}\label{sum}

We have proposed a scheme to realize  PT symmetry via EIT. The
system we considered is an ensemble of cold four-level atoms with
an EIT core. We have shown that the cross-phase modulation
contributed by an assisted field, the optical lattice potential
provided by a far-detuned laser field, and the optical gain coming
from an incoherent pumping can be used to construct a PT-symmetric
complex optical potential for probe field propagation in a
controllable way. Comparing with previous study in~\cite{hang},
our scheme has the following advantages: (i) Our scheme uses only
one atomic species, which is much simpler than that
in~\cite{hang}. (ii) The mechanism of realizing the PT-symmetric
potential is based on EIT, which is different from that
in~\cite{hang} where a double Raman resonance was used. (iii) One
can design many different PT-symmetric potentials at will in our
scheme in a simple way.

\section*{Appendix}

\subsection*{Explicit expression of Eq.~(\ref{bloche})}

Equations of motion for $\sigma_{ij}$ are given by
\bes
\label{den}
\bea
& & i\f{\pl}{\pl t}\sigma_{11}+i\Gamma_{31}\sigma_{11}-i\Gamma_{13}\sigma_{33}+\Omega_{p}^{\ast}
\sigma_{31}-\Omega_{p}\sigma_{31}^{\ast}=0,\label{den1}\\
& & i\f{\pl}{\pl
t}\sigma_{22}-i\Gamma_{23}\sigma_{33}-i\Gamma_{24}\sigma_{44}+\Omega_{c}^{\ast}
\sigma_{32}-\Omega_{c}\sigma_{32}^{\ast}+\Omega_{a}^{\ast}
\sigma_{42}-\Omega_{a}\sigma_{42}^{\ast}=0,\label{den2}\\
& & i\left( \f{\pl}{\pl t}+\Gamma_{3}\right)
\sigma_{33}-i\Gamma_{31}\sigma_{11}
    -\Omega_{p}^{\ast}
\sigma_{31}+\Omega_{p}\sigma_{31}^{\ast}-\Omega_{c}^{\ast}
\sigma_{32}+\Omega_{c}\sigma_{32}^{\ast}=0,\label{den3}\\
& & i\left( \f{\pl}{\pl t}+\Gamma_{4}\right) \sigma_{44}
    -\Omega_{a}^{\ast}
\sigma_{42}+\Omega_{a}\sigma_{42}^{\ast}=0,\label{den4}\\
& & \left(i\f{\pl}{\pl
t}+d_{21}\right)\sigma_{21}+\Omega_{c}^{\ast}\sigma_{31}
+\Omega_{a}^{\ast}\sigma_{41}-\Omega_{p}\sigma_{32}^{\ast}
=0,\label{den21}\\
& & \left(i\f{\pl}{\pl
t}+d_{31}\right)\sigma_{31}+\Omega_{p}(\sigma_{11}-\sigma_{33})
+\Omega_{c}\sigma_{21}
=0,\label{den31}\\
& & \left(i\f{\pl}{\pl
t}+d_{41}\right)\sigma_{41}+\Omega_{a}\sigma_{21}
-\Omega_{p}\sigma_{43}=0,\label{den41}\\
& & \left(i\f{\pl}{\pl
t}+d_{32}\right)\sigma_{32}+\Omega_{c}(\sigma_{22}-\sigma_{33})
+\Omega_{p}\sigma_{21}^{\ast}-\Omega_{a}\sigma_{43}^{\ast}
=0,\label{den32}\\
& & \left(i\f{\pl}{\pl
t}+d_{42}\right)\sigma_{42}+\Omega_{a}(\sigma_{22}-\sigma_{44})
-\Omega_{c}\sigma_{43}=0,\label{den42}\\
& & \left(i\f{\pl}{\pl
t}+d_{43}\right)\sigma_{43}+\Omega_{a}\sigma_{32}^{\ast}
-\Omega_{p}^{\ast}\sigma_{41}-\Omega_{c}^{\ast}\sigma_{42}
=0,\label{den43}
\eea
\ees
with
$d_{jl}= \Delta^{\prime}_{j}-\Delta^{\prime}_{l}+i\gamma_{jl}$
($\Delta^{\prime}_j$ is given by Eq.~(\ref{Det})\,),
$\gamma_{jl}= (\Gamma_{j}+\Gamma_{l})/2+\gamma^{{\rm
dph}}_{jl}$ ($j\neq 3, l\neq 1$),
$\gamma_{31}= (\Gamma_{3}+\Gamma_{31})/2+\gamma^{{\rm dph}}_{31}$,
and $\Gamma_l = \sum_{E_j<E_l}\Gamma_{jl}$.  Here
$\gamma_{jl}^{\text{dph}}$ denotes the dipole dephasing rates
caused by atomic collisions; $\Gamma_{jl}$ is the rate at which population decays
from $|l\rangle$ to $|j\rangle$. Especially, $\Gamma_{31}$
is the incoherent pumping rate from $|1\rangle$ to  $|3\rangle$.

\subsection*{Perturbation expansion of the MB equations}

The coefficients of Eq.~(\ref{Vt}) are given by
\bes
\bea
& &\label{a112} \alpha_{12}=-\f{\kappa_{13}\Omega_{c}}{D_{1}}\alpha_{41}^{(2)}
+\f{\kappa_{13}\Omega_{c}}{D_{1}}\alpha_{23G}^{(2)}+\f{\kappa_{13}d_{21}^{(0)}}{D_{1}}
(\alpha_{11G}^{(2)}-\alpha_{33G}^{(2)}),\\
& & \label{al13} \alpha_{13} =
\f{\kappa_{13}(\alpha_{3}-\alpha_{1})}{2\hbar
D_{1}}d_{21}^{(0)}\alpha_{31}^{(1)},
\eea
\ees
where
\bes
\bea
& & \alpha_{22F}^{(2)}=\f{-2\Gamma{\rm
Im}\left[\f{\Omega_{c}^{\ast}(\alpha_{21}^{(1)})^{\ast}}{d_{32}^{(0)}}\right]-2(\Gamma_{23}-X_{3}){\rm
Im}(\alpha_{31}^{(1)})}{\Gamma
X_{3}-\Gamma_{31}(\Gamma_{23}-X_{3})}, \\
& & \alpha_{33F}^{(2)}=\left[2{\rm
Im}(\alpha_{31}^{(1)})-\Gamma_{31}\alpha_{22F}^{(2)}\right]/\Gamma,\\
& & \alpha_{22G}^{(2)}=\f{2\Gamma{\rm
Im}[\f{\Omega_{c}^{\ast}(\alpha_{43}^{(1)})^{\ast}}{d_{32}^{(0)}}]+2\Gamma
{\rm Im}(\alpha_{42}^{(1)})+2\Gamma_{31}(\Gamma_{23}-X_{3}){\rm
Im}(\alpha_{42}^{(1)})/\Gamma_{4}}{\Gamma
X_{3}-\Gamma_{31}(\Gamma_{23}-X_{3})}, \\
& & \alpha_{44}^{(2)}=\f{2}{\Gamma_{4}}{\rm
Im}(\alpha_{42}^{(1)}),\\
& & \alpha_{41}^{(2)}=\f{\alpha_{43}^{(1)}-\alpha_{21}^{(1)}}{d_{41}^{(0)}},\\
& & \alpha_{33G}^{(2)}=(-\Gamma_{31}\alpha_{44}^{(2)}-\Gamma_{31}\alpha_{22G}^{(2)})/\Gamma,\\
& & \alpha_{11F}^{(2)}=[\Gamma_{13}\alpha_{33F}^{(2)}-2{\rm
Im}(\alpha_{31}^{(1)})]/\Gamma_{31},\\
& &  \alpha_{11G}^{(2)}=\Gamma_{13}\alpha_{33G}^{(2)}/\Gamma_{31},\\
& &  \alpha_{23F}^{(2)}=(-\alpha_{21}^{(1)}+\Omega_{c}^{\ast}\alpha_{33F}^{(2)}-\Omega_{c}^{\ast}
\alpha_{22F}^{(2)})/(d_{32}^{(0)})^{\ast},\\
& & \alpha_{23G}^{(2)}=(\alpha_{43}^{(1)}+\Omega_{c}^{\ast}\alpha_{33G}^{(2)}
-\Omega_{c}^{\ast}\alpha_{22G}^{(2)})/(d_{32}^{(0)})^{\ast},
\eea
\ees
with $\Gamma=\Gamma_{13}+\Gamma_{31}$.


\section*{Acknowledgments}

This work was supported by the NSF-China under Grant Nos.
11074221, 11174080, and 11204274, and by the discipline
construction funds of ZJNU under Grant No. ZC323007110; and in
part by the Open Fund from the State Key Laboratory of Precision
Spectroscopy, ECNU.

\end{document}